\documentstyle[aps,prl,multicol,epsf,amssymb]{revtex}
\def\gs{\gtrsim}
\def\ls{\lesssim}
\def\gdot{\dot{\gamma}}
\def\be{\begin{equation}}
\def\en{\end{equation}}                  
\newcommand{\bi}[1]{\mbox{\boldmath$#1$}}
\newcommand{\av}[1]{\langle{#1}\rangle}

\begin{document}
\draft
\bibliographystyle{prsty}
\title{Entanglements in a Quiescent and Sheared Polymer Melt}
\author{Ryoichi Yamamoto and Akira Onuki}
\address{Department of Physics, Kyoto University, Kyoto 606-8502, Japan}
\date{\today}
\maketitle

\begin{abstract}
Molecular dynamics simulations were performed for a polymer melt.  
In quiescent states 
the inter-chain interaction energy supported by each particle 
takes relatively large values persistently for long times 
if the particle is close to an entanglement. 
Thus, if  the interaction is averaged over appropriate time intervals, 
we  can detect  active spots  on the chains 
produced by entanglement constrains.  If rapid shearing is  applied, 
these active regions subsequently 
 become bended in eventual zig-zag shapes 
of the chains. 
We also demonstrate 
 that stress overshoot occurs with onset of 
disentanglement. 
\end{abstract}

\pacs{PACS numbers: 83.10.Nn, 83.20.Jp, 83.50.By, 61.25.Hq}

\begin{multicols}{2}
\narrowtext

%Introduction

The dynamics of dense polymer melts 
has been a challenging subject 
in current polymer physics 
\cite{degennes,doi}.
While the dynamics of short chains can be reasonably 
described by the simple Rouse model, the dynamics of very long chains 
has not yet been well 
understood on the microscopic level since it is 
governed by  entanglement effects.
The reptation theory \cite{degennes,doi} is the most successful 
 approach to date in describing the dynamics 
of entangled polymer chains. 
It has been supported by 
 a number of experimental  
and numerical works \cite{schleger,kremer,smith,ebert,paul,putz,hess,aoyagi}, 
where   experimentally accessible quantities such as 
the stress relaxation function $G(t)$, 
the incoherent dynamic scattering function,
and the mean square displacement  have been 
compared with the theoretical predictions.
Understanding rheological 
 properties  of polymer melts is 
 also of great importance and 
extensive simulations have been performed in this direction 
\cite{hess,aoyagi,yo1}.
In a unique paper, 
Ben-Naim {\it et al.} \cite{ben-naim} 
could visualize
 individual entanglements using 
 the fact that incidental contacts of the particles 
and entanglement contacts behave differently 
because entanglement constraints are long-lived. 
The main motivation of our simulation is to identify 
entanglements more 
systematically 
 on the microscopic level and examine how they 
influence  rheology in rapid shearing.

%Model 
We used the bead-spring model \cite{kremer} for our polymer melt 
composed of  $M$ chains with $N$ beads. 
All the bead particles interact via the Lennard-Jones potential,
$U_{\rm LJ}(r)= 4\epsilon [(\sigma/r)^{12}-(\sigma/r)^{6}] + \epsilon$, 
truncated at the minimum distance $2^{1/6}\sigma$. 
It serves to prevent spatial overlap of the particles. 
In addition, bonded pairs or consecutive beads on each chain  
are connected by an anharmonic spring potential of the form, 
$U_{\rm FENE}(r)= -\frac{1}{2}k_c R_0^2 \ln[1-(r/R_0)^2]$ 
with $k_c=30\epsilon/\sigma^2$ and $R_0=1.5\sigma$. As a result, 
the bond lengths  deviate within only a few $\%$ from the minimum 
of $U_{\rm LJ}(r)+U_{\rm FENE}(r)$ given by $0.96\sigma$. 
The number density and temperature were fixed at 
$n= NM/V =1/\sigma^3$ and $T=\epsilon/k_B$. 
We measure space and time in units of 
$\sigma$ and $\tau_0=({m\sigma^{2}/\epsilon})^{1/2}$ with 
$m$ being the bead mass. 
We numerically solved the Newton's equation of motion
and took data after long equilibration periods 
of order $10^6$ to avoid aging (slow equilibration).  
In quiescent cases, we imposed the micro-canonical condition with 
the time step $\Delta t = 0.005$ for numerical integration. 
In order to obtain accurate linear viscoelastic behavior of entangled 
polymers, very long simulations of order $10 \tau_d$ were 
performed for a system composed of $M=10$ chains each
consisting of $N=250$ beads, where $\tau_d=6\times10^5$ is 
the stress  relaxation time.
In the presence of shear flow, we set $\Delta t=0.0025$ and kept   
the temperature at a constant using the Gaussian constraint 
thermostat to eliminate viscous heating.
After a long equilibration time in a quiescent state for $t < 0$, 
all the particles acquired a velocity $\gdot y$ 
in the $x$ direction at $t = 0$, 
and then the Lee-Edwards boundary 
condition \cite{Allen,Evans} maintained the simple shear flow for $t > 0$.
Steady sheared states were realized after transient nonlinear 
viscoelastic regime.

%Results
%%%%%% quiescent 
First, we show data of the stress relaxation function,
\be
G(t) = \langle P_{xy}(t)P_{xy}(0)\rangle/V k_BT,
\label{eq:1}
\en 
at zero shear rate in Fig.1, where $P_{xy}$ is the $xy$ component of 
the total stress tensor.  
For the shorter chain case of $N=25$ and $M=40$, 
$G(t)$ is well fitted to the Rouse relaxation function \cite{verdier}, 
$
G_R(t)=({n k_B T}/{N}) 
\sum_{p=1}^{N-1}
\exp(-t/\tau_p),
$ 
where $\tau_p=\tau_{01} /\sin^2({p\pi}/{2N})$ 
with $\tau_{01}=\zeta b^2/12k_BT\cong 6.5$ \cite{zeta}.
$G(t)$ for the longer chain case of 
$N=250$ and $M=10$ was obtained from a single extremely long
simulation ($\sim 5\times10^6=10^9\Delta t$). 
It clearly exhibits slow relaxation due to entanglements.  
Indeed, it can reasonably well be fitted to the stress relaxation 
function from the reptation theory,
$
G_{rep} (t)=({8n k_B T}/{\pi^2 N_e}) 
\sum_{p= {\rm odd} }
\exp(-p^2t/\tau_d)/{p^2},
$
with $N=250$, $N_e=100$, and $\tau_d=6\times10^5$ for 
$t\gs 2\times 10^5$. 
 
In order to identify and visualize entanglements for the case 
$N=250$, we define the potential energy of non-bonded interaction 
on the $i$-th particle by 
\be 
E^{\rm NB}_i(t)= \sum_{j \in {\rm non~ bond}}
U_{LJ} (|{\bi r}_i(t)-{\bi r}_j(t)|) , 
\label{eq:2}
\en 
where the particle $j$ is not bonded to  $i$. 
The distribution of $E^{\rm NB}_i(t)$ at each time $t$ 
is nearly Gaussian, and there is no appreciable correlation even 
between adjacent beads, 
 because $E^{\rm NB}_i(t)$ consists mostly of 
rapidly varying thermal fluctuations.  To eliminate the rapid components 
 we examine its time average, 
\be
\bar{E}^{\rm NB}_i(t)=\frac{1}{\tau} 
\int_0^{\tau} dt'  {E}^{\rm NB}_i (t+t'), 
\label{eq:3}
\en 
where the time interval $\tau$ is much longer than 1. 
We find that the variance,  
\be 
\sigma(\tau) = \sqrt{ \frac{1}{MN}
\sum_i  \bigg (\bar{E}^{\rm NB}_i- \av{\bar{E}^{\rm NB}} \bigg )^2 }, 
\label{eq:4}
\en 
decreases as $\tau^{-1/2}$ with increasing $\tau$ in the region 
$ 1\ll \tau \ll \tau_d$, where $\av{\bar{E}^{\rm NB}}$ is the average. 
This is  because the rapidly varying contributions behave as white-noises.  
With this time-averaging procedure, strong correlations due to entanglements 
become apparent in the non-bonded interactions along the chains.  
Fig.2 displays instantaneous values ${E}^{\rm NB}_i$ in (a) and 
time-averaged values $\bar{E}^{\rm NB}_i$ with $\tau= 5\times 10^3$ in (b). 
The chains are straightened horizontally for the visualization purpose. 
While no correlations can be seen in (a),  we can see 
active spots which consist of several 
consecutive beads having 
distinctly large  values of 
$\bar{E}^{\rm NB}_i$. Then we recognize that 
there are  two or three entanglements on each chain.  
Furthermore, to examine the correlations in $\bar{E}^{\rm NB}_i$
along the chains quantitatively, we define the auto-correlation function,
\be
C(n)=\frac{1}{\cal{N} } 
\sum_{j-i=n} \bigg 
 (\bar{E}^{\rm NB}_i- \av{\bar{E}^{\rm NB}}\bigg )
\bigg (\bar{E}^{\rm NB}_j- \av{\bar{E}^{\rm NB}} \bigg ), 
\label{eq:5}
\en
where the two beads $i$ and $j$ are separated by $n$ on the same chain. 
The normalization factor $\cal{N}$ is defined such that $C(0)=1$; 
then, ${\cal{N}} = MN  \sigma(\tau)^2$.  
In Fig.3, we show $C(n)$ in (a) and its power spectrum, 
\be
P(k)=\sum_{n=0}^{N-1}C(n)\exp ( -i k n) ,
\label{eq:6}
\en
in (b) for various $\tau$ in the range $0 \le \tau \ll \tau_d$. 
The correlation between adjacent beads $C(1)$ is nearly zero for 
$\tau=0$ but increases with increasing $\tau$.
The  decay of $C(n)$ for $n \ls 10$ 
 becomes slower  with increasing $\tau$. 
The width $\Delta n$ determined by 
$C(\Delta n) \ls  10^{-2}$  in Fig.3a grows 
with increasing $\tau$ roughly as 
$\Delta n \propto \tau^{a}$ with $a \simeq 0.35$.  
This widening  arises from the 
 motion of  entanglements along the chains 
or equivalently from the motion of chains through 
tubes  formed by 
entanglement constraints \cite{degennes,doi}. 
At $\tau=5\times 10^3$,  C(n) takes a  negative 
minimum around $n \simeq 45$ 
and a positive maximum around $n \simeq 80$. 
Correspondingly, $P(k)$ has a peak at $k  \simeq 2\pi/80$.  
Thus the average bead number $N_e$ between 
entanglements is about 80, which 
is consistent with recent  simulation results \cite{putz}.
For much larger $\tau$ ($ >5\times10^{4}$), however, 
no periodic structure of $C(n)$ was  observed because 
entanglements become delocalized along the chains 
on such very long time scales.

%%%%%%%% shear 
Next, we applied a  shear flow with  rate 
$\gdot=10^{-3}(\simeq 500/\tau_d)$  using the same 
initial condition which gave the data shown in Fig.2b. 
In Fig.4 the chain conformations are projected 
on the $xz$ plane at $\gdot t=5$ in (a) and $\gdot t=10$ in (b). 
Because the chains are rapidly elongated, 
they eventually take zig-zag shapes 
bended at entanglements. 
 Remarkably,  the  active  regions detected without shear   in Fig.2b 
mostly become bended under shear in Fig.4a. 
Here we give the same number to 
an active spot in Fig.2a and that in Fig.4a on the same 
 chain if their 
distance measured 
along the chain remains shorter  than 10. 
This coincidence unambiguously demonstrates the validity of 
the time-averaging method used in Fig.2b to visualize 
entanglements in quiescent states.
Furthermore, the non-bonded interactions in these entangled  regions 
(mostly in bended parts) become increasingly amplified with 
increasing the strain, obviously because 
a considerable fraction of the stress is supported by 
 entanglements in strong deformations.  
As a result, they can be detected even with  much smaller $\tau$ 
than in the quiescent condition, so we set $\tau=500$ in 
calculating $\bar{E}^{\rm NB}_i$ in  Fig.4. 
From Fig.4a to Fig.4b the 
 movement of the entangled  regions could be followed, 
so the corresponding 
parts in Fig.4a and Fig.4b  can be  marked with  
 the same number. We can also see that the total number of these regions 
has not yet decreased in Fig.4b at $\gdot t=10$, but 
several of them are approaching   chain ends and are 
 going to disappear.  Note that entanglements can be released 
only when they reach a chain end.    Thus, the 
disentanglement process induced by shear flow 
is going to start in Fig.4b.

In the case of entangled melts, the shear stress $P_{xy}(t)$ 
and the normal stress $P_{xx}(t)-P_{yy}(t)$ often exhibit
overshoot behavior 
in  rapid shearing \cite{hess,aoyagi,mead}. 
In Fig.5 we show the shear stress $P_{xy}(t)$ 
and the normal stress difference $P_{xx}(t)-P_{yy}(t)$ after application of 
shear at $\gdot=10^{-3}$. They were 
 calculated in the same run yielding Fig.4. 
In agreement with experiments and the previous simulations, 
(i) these stress components both exhibit 
a peak around $\gdot t \simeq 10$ 
and afterwards tend to a steady state value 
and (ii) the peak  of  $P_{xx}(t)-P_{yy}(t)$ is attained shortly 
after that of $P_{xy}(t)$.
As can be seen in Fig.4b, the chains are stretched without 
appreciable disentanglement until the stress maximum.  
We recognize  that the stress components begin to decrease 
with onset of the disntanglement process.

%Summary
In summary, we have succeeded to detect and 
visualize the entanglements in a model polymer melt 
in both  quiescent and sheared conditions to obtain  $N_e \simeq 80$.
%Acknowledgement
This work is supported by Grants in Aid for Scientific 
Research from the Ministry of Education, Science, Sports and Culture
of Japan.
Calculations have been performed at the Human Genome Center, 
Institute of Medical Science, University of Tokyo and 
the Supercomputer Center, Institute for Solid State Physics, 
University of Tokyo.

\end{multicols}

\narrowtext
%%%%% Fig.1 %%%%%%%%%%%%%%%%%%
\begin{figure}[t]
\centerline{\epsfxsize=2.8in\epsfbox{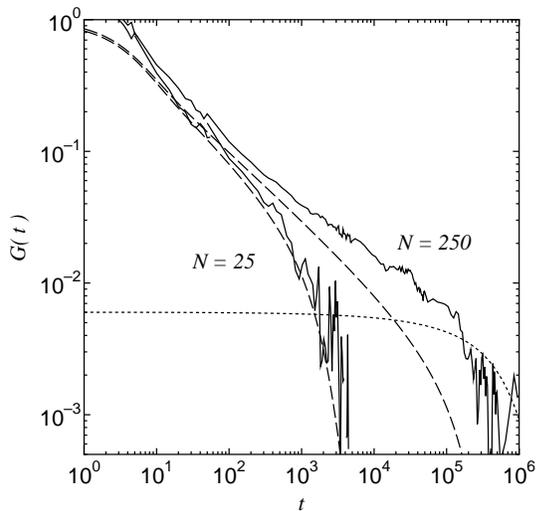}}
\caption{\protect%\narrowtext 
The calculated 
stress relaxation function $G(t)$ 
 for $N=25$ and $250$ (solid lines). 
The  Rouse relaxation function 
$G_R(t)$ (doted lines) and 
that from the reptation theory 
$G_{rep}(t)$ (dashed line)
with $N_e=100$ are also shown.
}
\label{fig1}
\end{figure}

\newpage
\widetext
%%%%%  Fig.2 %%%%%%%%%%%%%%%%%%
\begin{figure}[b]
\noindent
\centerline{\epsfxsize=3.5in\epsfysize=1.8in\epsfbox{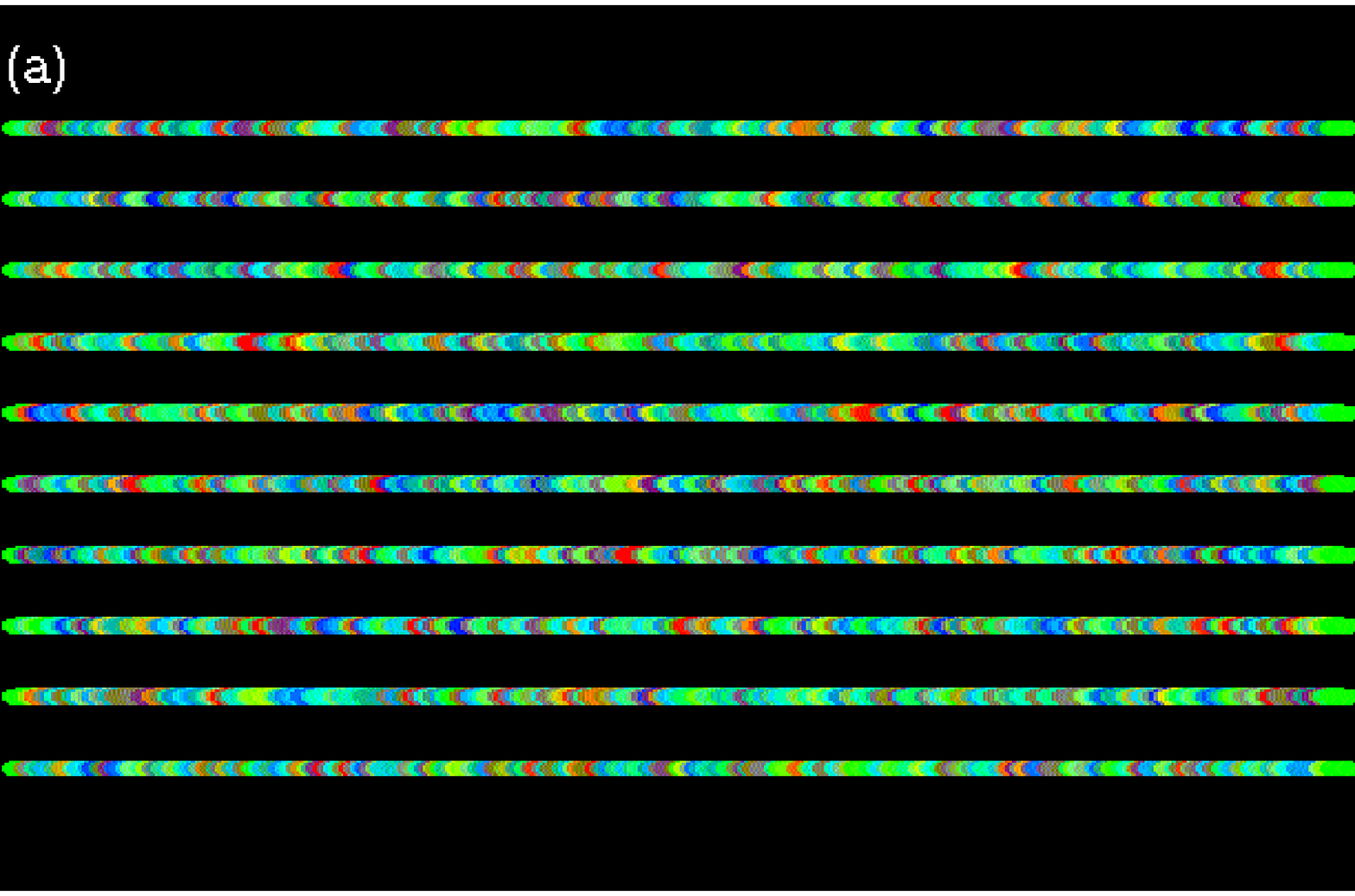}
\epsfxsize=3.5in\epsfysize=1.8in\epsfbox{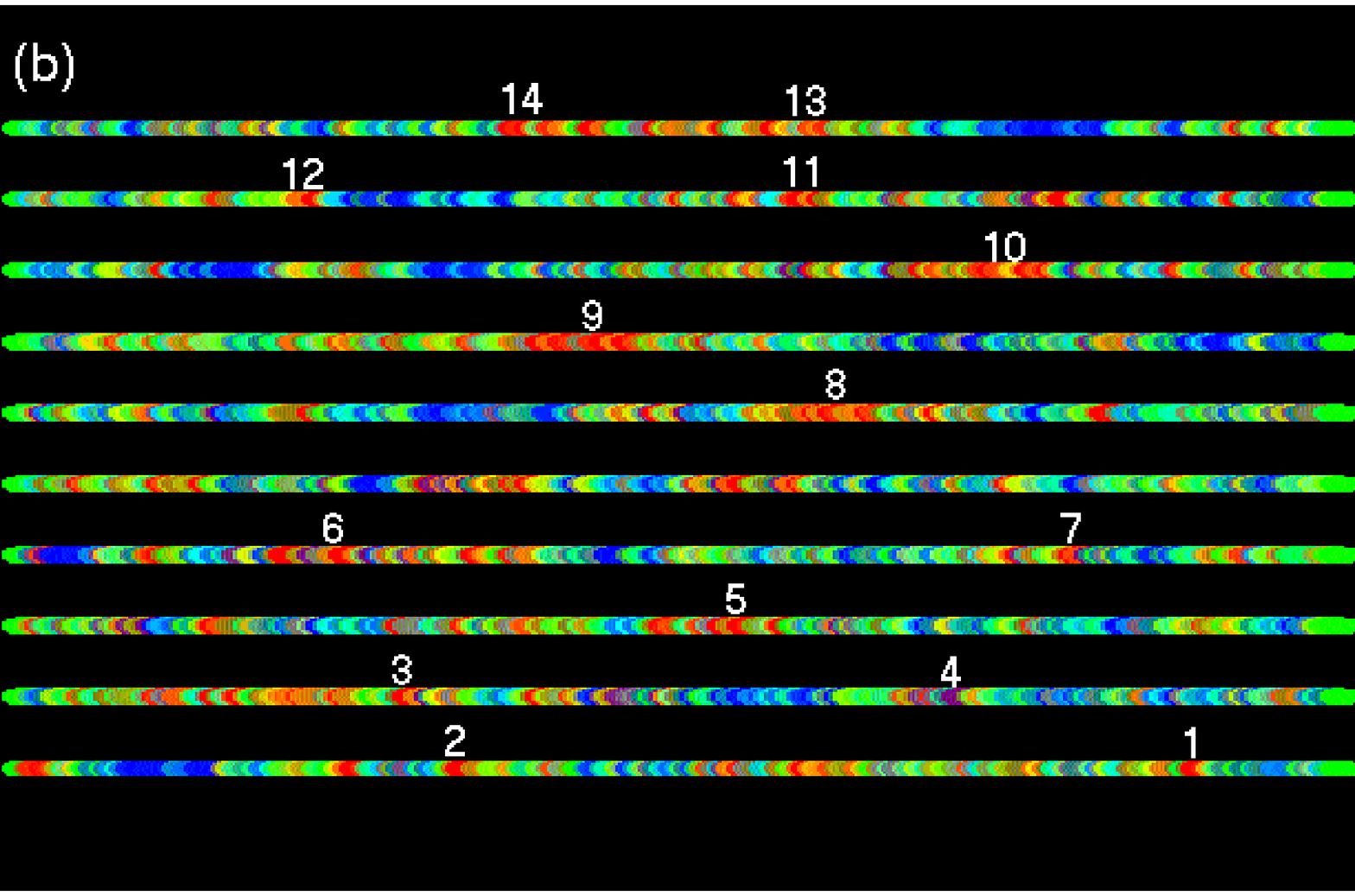}}
\vspace{5mm}
\caption{\protect%\narrowtext
Distributions of non-bonded interactions on the chains with 
$N=250$ in a quiescent state. 
We can see no correlations along the chains 
 in snapshot values of  
$[E^{\rm NB}_i(t)- \av{E^{\rm NB}}] /\sigma (0)$ in 
(a).  However, the time-averaged 
 normalized  interactions 
$[\bar{E}^{\rm NB}_i(t)- 
\av{E^{\rm NB}}] /\sigma (\tau)$ with  $\tau = 
5 \times 10^3$  are distinctly strong in line segments  
 consisting of several beads (in red or orange) due to entanglements. 
Among these segments 
we number those which keep to hold large  non-bonded interaction 
values during  rapid shearing in Fig.4.  
The  color map is the same as  in Fig.4. 
}
\label{fig2}
\end{figure}

%\newpage
\widetext
%%%%%  Fig.3 %%%%%%%%%%%%%%%%%%
\begin{figure}[t]
\centerline{\epsfxsize=2.8in\epsfbox{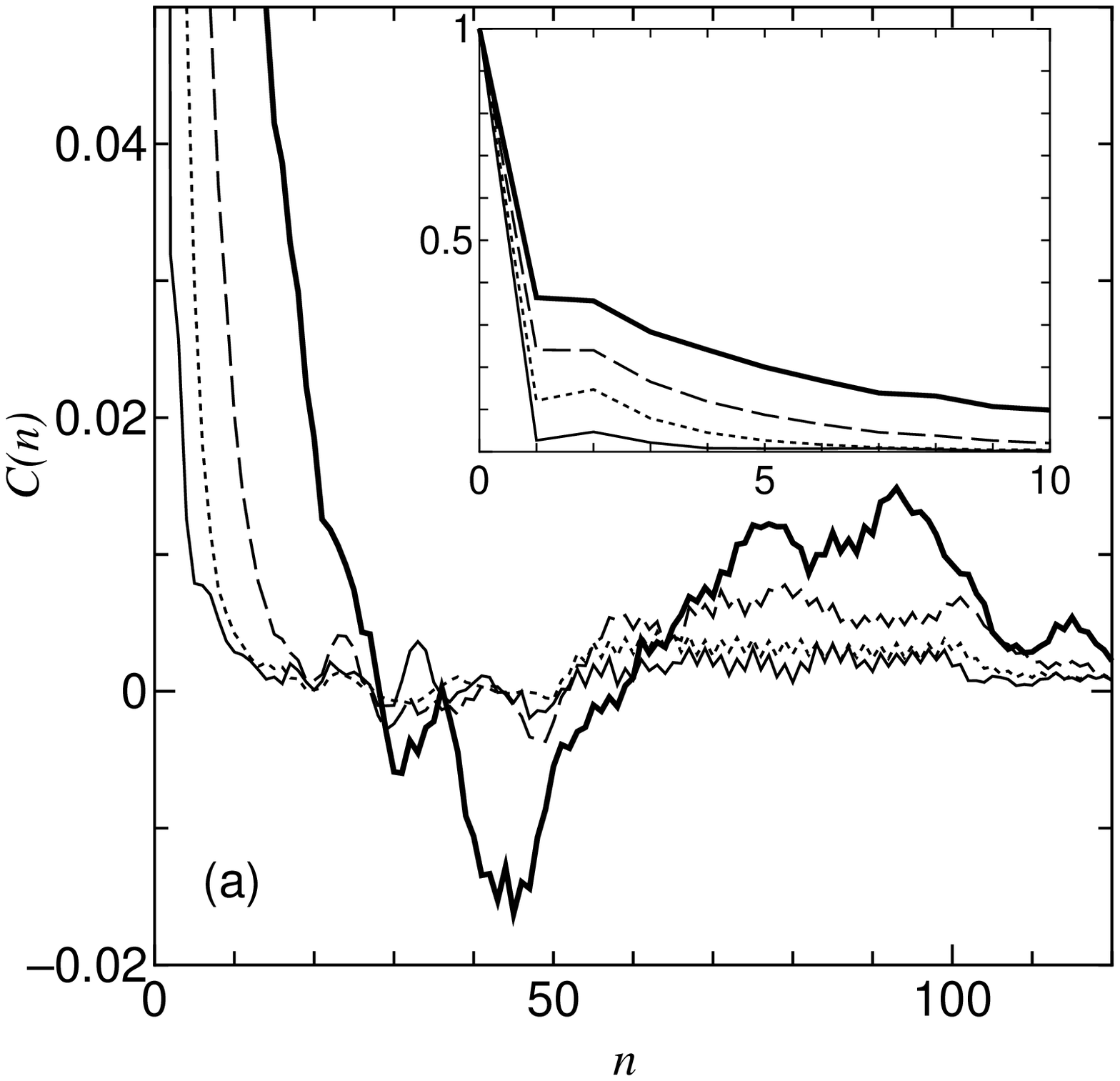}\hspace{10mm}
\epsfxsize=2.8in\epsfbox{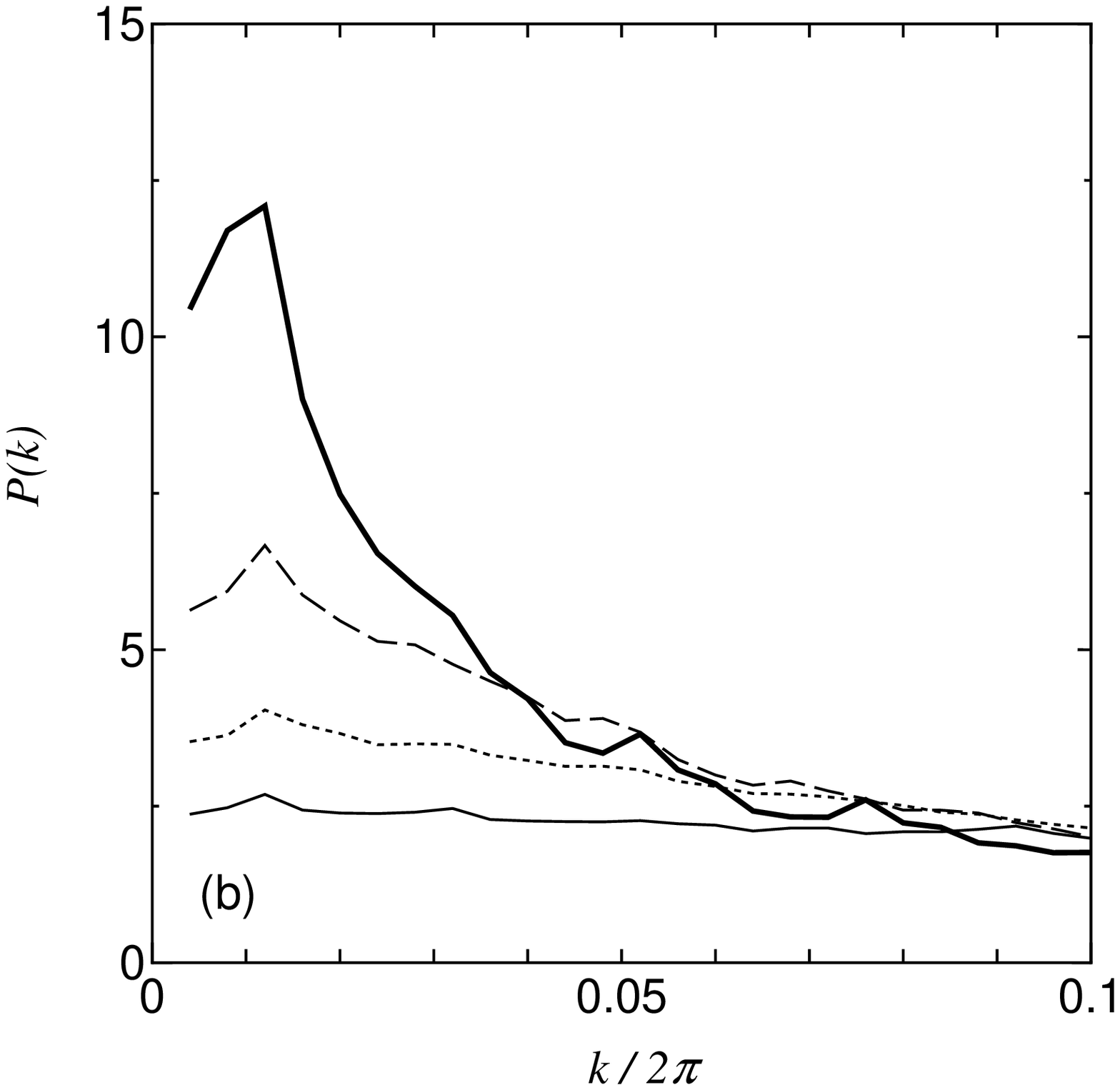}}
\caption{\protect%\narrowtext 
The auto-correlation function $C(n)$ defined 
by Eq.(5) in  (a),  and its power spectrum  $P(k)$ defined by 
Eq.(6) in (b), for various values of the time 
interval $\tau$.  
The inset in (a)  shows the decay of $C(n) $ 
for small $n$. 
Here $\tau=0$ (thin-solid line), 
 $5\times10$ (dotted line), $5\times10^2$ (dashed line), 
and $5\times10^3$ (bold line).
}
\label{fig3}
\end{figure}

\newpage
\widetext
%%%%%  Fig.4 %%%%%%%%%%%%%%%%%%
\begin{figure}[t]
\centerline{\epsfxsize=3.5in\epsfbox{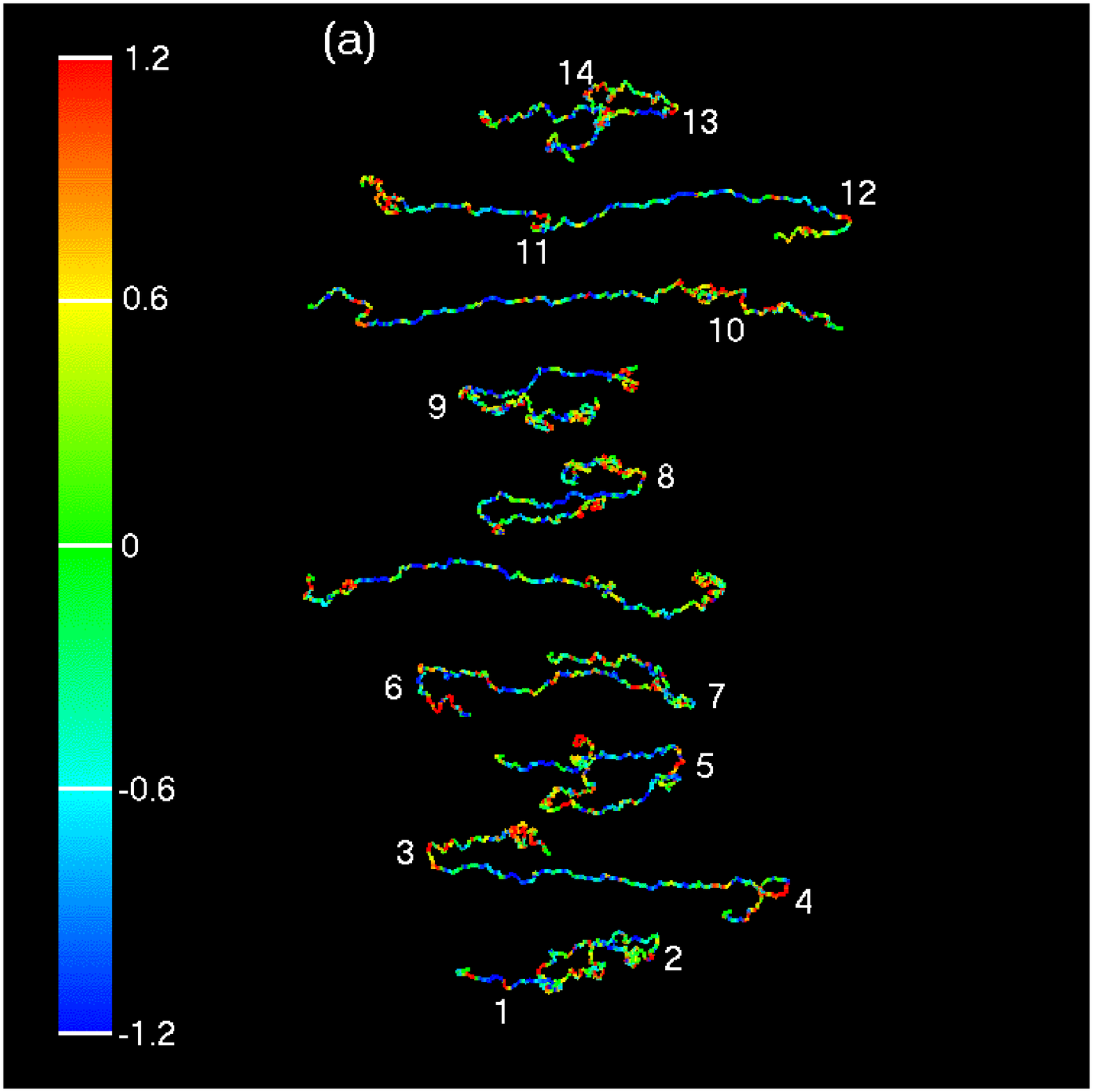}
\epsfxsize=3.5in\epsfbox{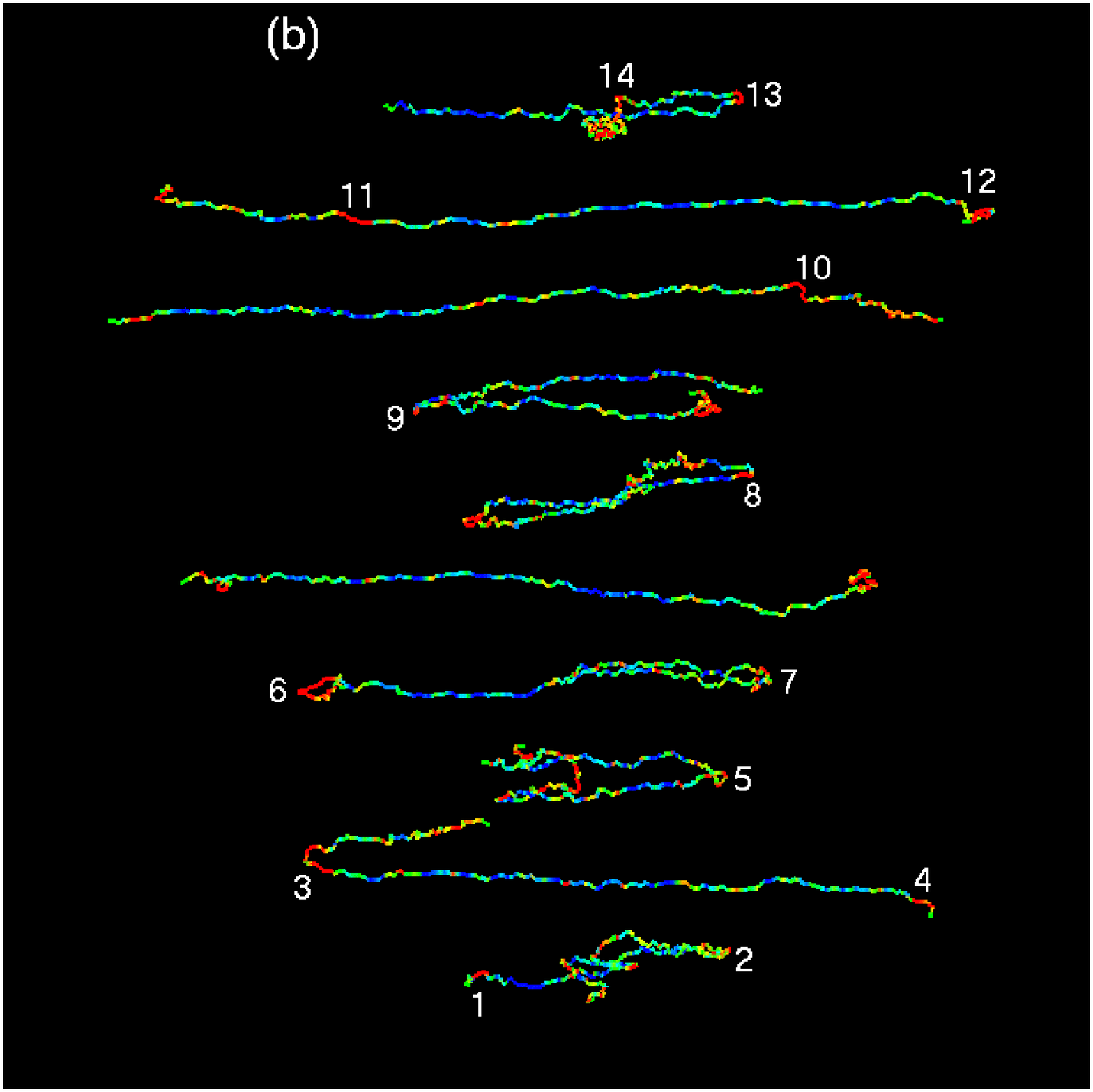}}
\vspace{5mm}
\caption{\protect%\narrowtext
Snapshots of the deformed chains in the $xz$ plane. 
A shear flow 
 at $\gdot=10^{-3}$ is applied to the system with 
the configuration in Fig.2b in the same run. 
 The numbered 
segments correspond to those in Fig.2b. 
Here $\gdot t = 5$ in (a) and 
 $\gdot t = 10$ in (b).  The flow is 
in the horizontal ($x$) direction, and the shear gradient  is 
 in the out-of-plane ($y$) direction. 
The non-bonded interactions 
with $\tau=500$ are written on the chains.  
Use is made of the  color map  on the left, where the 
numbers are the normalized deviations,
$(\bar{E}^{NB}_i-\av{\bar{E}^{NB}}) /\sigma(\tau)$, with $\tau=500$.  
}
\label{fig4}
\end{figure}

%\newpage
\narrowtext
%%%%%  Fig.5 %%%%%%%%%%%%%%%%%%
\begin{figure}[t]
\centerline{\epsfxsize=2.8in\epsfbox{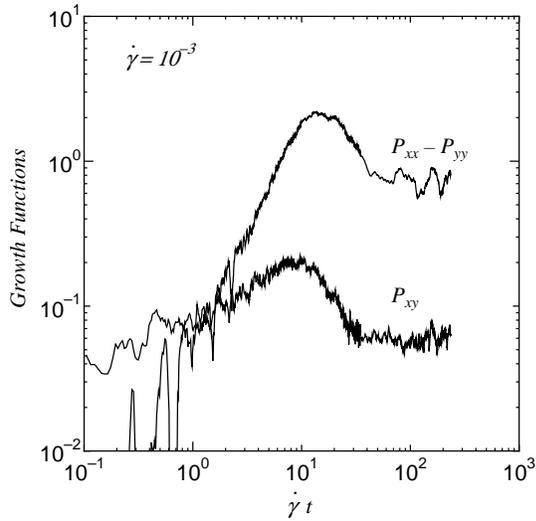}}
\caption{\protect%\narrowtext 
The stress growth functions after application of shear flow 
at  $\gdot=10^{-3}$ in a model polymer melt with $N=250$. 
The  chain shapes and entanglements 
at  $\gdot t = 10$  are written in Fig.4b.
}
\label{fig5}
\end{figure}


\begin{references}
\bibitem{zeta} We also calculated 
the time-correlation function of the end-to-end vector 
$\av{{\bi P}(t)\cdot {\bi P}(0)}$  and the incoherent 
 dynamic scattering function  $S_{inc}(q,t)$. 
 From these data  we obtained 
$\zeta/k_BT \cong 52\tau_0/\sigma^2$ 
and $b^2\cong 1.5 \sigma^2$  for $10\le N\le250$.
\end{references}
\end{document}